\documentclass[prd,nofootinbib,showpacs,preprint]{revtex4}

\usepackage{graphicx}
\usepackage[usenames,dvipsnames]{color}
\usepackage{amsmath,amssymb}
\usepackage[colorlinks]{hyperref}
\begin{document}
\title{Abelian Gauge Extension of Standard Model: Dark Matter and Radiative Neutrino Mass}
\author{Debasish Borah}
\email{debasish@phy.iitb.ac.in}
\affiliation{Indian Institute of Technology Bombay, Mumbai 400076, India}
\author{Rathin Adhikari}
\email{rathin@ctp-jamia.res.in}
\affiliation{Centre for Theoretical Physics, Jamia Millia Islamia - Central University, Jamia Nagar, New Delhi 110025, India}
\begin{abstract}
We study a simple extension of Standard Model where the gauge group is extended by an additional $U(1)_X$ gauge symmetry. Neutrino mass arise both at tree level as well as radiatively by the anomaly free addition of one singlet fermion $N_R$ and two triplet fermions $\Sigma_{1R}, \Sigma_{2R}$ with suitable Higgs scalars. The spontaneous gauge symmetry breaking is achieved in such a way which results in a residual $Z_2$ symmetry and hence providing a stable cold dark matter candidate. We study the possible dark matter candidates in this model by incorporating the constraints from cosmology as well as direct detection experiments. We discuss both low and high mass (from GeV to the TeV scale) regimes of fermionic and scalar dark matter candidates in the model. We show that scalar dark matter relic density, although not significantly affected by the presence or absence of annihilation into $U(1)_X$ gauge boson pairs, is however affected by choice of $U(1)_X$ gauge charges. We discuss the neutrino mass phenomenology and its compatibility with the allowed dark matter mass ranges and also comment on the implications of the model on Higgs signatures at colliders including those related to fourth fermion generation.
\end{abstract}
\maketitle

\section{Introduction}

The Standard Model (SM) of particle physics has been phenomenologically the most successful low energy effective theory for the last few decades. The 
predictions of standard model have been verified experimentally 
with a very high accuracy except the missing Higgs boson. Despite its phenomenological success, we all now know that this model neither address 
many theoretical issues like 
gauge hierarchy problem, nor provides a complete understanding of various observed phenomena like non-zero neutrino masses, dark matter etc. 
A great deal of works have been done so far on various possible extensions of the Standard model, although none of them can be called a complete 
phenomenological model. Such extensions generally involve incorporating some extra symmetries into the Standard model or inclusion of additional fields. We know that the smallness of three Standard Model
neutrino masses \cite{neutosc} can be naturally explained 
via seesaw mechanism. Such seesaw mechanism can be of three types : type I \cite{typeI}, type II \cite{typeII} and type III \cite{Foot:1988aq}. All these mechanisms involve the inclusion of additional fermionic or scalar fields to generate tiny neutrino masses at tree level. However, it may be true that the gauge structure and the particle content of the theory do not allow neutrino masses at tree level and a tiny neutrino mass appears only at the loop level. Here we are interested in a model which gives rise to such radiative neutrino mass in the manner proposed in \cite{Ma:2006km,Adhikari:2008uc}.

In addition to extra scalar and fermionic fields, we also have an extended gauge structure in this model. The Standard Model gauge group is extended by an additional $U(1)_X$ gauge symmetry. It is worth mentioning that abelian gauge extension of Standard Model is one of the best motivating examples of beyond Standard Model physics \cite{Langacker:2008yv}. Such a model is also motivated within the framework of GUT models, for example $E_6$. 
The supersymmetric version of such models have an additional advantage in the sense that they provide a solution to the MSSM $\mu$ problem. An abelian gauge extension of SM was studied recently by one of us in the context of four fermion generations \cite{Borah:2011ve} which explains the origin of three light and one heavy fourth generation neutrino masses and at the same time provides a way to avoid the strict bounds put by Large Hadron Collider (LHC) on a SM like Higgs boson mass in the presence of a fourth family.

In the model considered here, the $U(1)_X$ gauge charges of different fields are chosen in such an anomaly free way that it allows neutrino mass only at one loop level and also gives rise to a remnant $Z_2$ symmetry so that the lightest $Z_2$-odd particle is stable and hence can be a cold dark matter candidate. Thus, the model provides a natural origin of the $Z_2$ symmetry which keeps the dark matter stable unlike in the generic supersymmetric models like MSSM where an ad-hoc $Z_2$ symmetry called R-parity is invoked to keep the dark matter stable as well as to keep the baryon and lepton number violating terms away from the Lagrangian. Motivated by the presence of such a natural $Z_2$ symmetry, we study all the electrically neutral $Z_2$-odd particles in our considered model in the context of dark matter by incorporating necessary constraints on relic density from Wilkinson Mass Anisotropy Probe (WMAP) data \cite{Jarosik:2010iu}. We also take into account the latest upper bounds on dark matter nucleon scattering cross sections from XENON experiment \cite{Angle:2008we,Aprile:2010um}. We also discuss the compatibility of our dark matter results with the observed neutrino mass scale and briefly comment on the LHC signatures in view of the recent LHC Higgs mass exclusion results \cite{lhc:2011,Koryton2011}. We note that a recent study of radiative neutrino mass and dark matter in an abelian extension of SM was carried out in \cite{takehiro}. However, our work is different in the sense that we consider a more general $U(1)_X$ gauge group rather than $U(1)_{B-L}$ in \cite{takehiro} and the radiative origin of neutrino mass considered here is entirely different from the one studied in \cite{takehiro}.

\indent This paper is organized as follows. In the next section \ref{model} we discuss the model in details: the gauge structure, the particle content and spontaneous symmetry breaking. We discuss fermionic dark matter candidates in section \ref{fermDM} and the scalar dark matter candidates in \ref{scalDM}. We then discuss the phenomenology of neutrino mass and its compatibility with the allowed dark matter mass range in section \ref{sec:numass} and also briefly comment on collider signatures in view of the recent LHC results in section \ref{sec:lhc}. We finally discuss our results and conclude in section \ref{results}.

\section{The Model}
\label{model}
The model we are going to discuss in detail was  proposed in \cite{Adhikari:2008uc}. The authors in that paper discussed various possible scenarios with different combinations of Majorana singlet fermions $N_R$ and Majorana triplet fermions $\Sigma_R$. Here we discuss one of such models which we find the most interesting as it naturally gives rise to a stable cold dark matter candidate and radiative origin of neutrino masses. This, so called model C by the authors in \cite{Adhikari:2008uc}, has the the following particle content shown in table \ref{table1}.

\begin{center}
\begin{table}
\caption{Particle Content of the Model}
\begin{tabular}{|c|c|c|c|}
\hline
Particle & $SU(3)_c \times SU(2)_L \times U(1)_Y$ & $U(1)_X$ & $Z_2$ \\
\hline
$ (u,d)_L $ & $(3,2,\frac{1}{6})$ & $n_1$ & + \\
$ u_R $ & $(\bar{3},1,\frac{2}{3})$ & $\frac{1}{4}(7 n_1 -3 n_4)$ & + \\
$ d_R $ & $(\bar{3},1,-\frac{1}{3})$ & $\frac{1}{4} (n_1 +3 n_4)$ & +\\
$ (\nu, e)_L $ & $(1,2,-\frac{1}{2})$ & $n_4$ & + \\
$e_R$ & $(1,1,-1)$ & $\frac{1}{4} (-9 n_1 +5 n_4)$ & + \\
\hline
$N_R$ & $(1,1,0)$ & $\frac{3}{8}(3n_1+n_4)$ & - \\
$\Sigma_{1R,2R} $ & $(1,3,0)$ & $\frac{3}{8}(3n_1+n_4)$ & - \\
$ S_{1R}$ & $(1,1,0)$ & $\frac{1}{4}(3n_1+n_4)$ & + \\
$ S_{2R}$ & $(1,1,0)$ & $-\frac{5}{8}(3n_1+n_4)$ & - \\
\hline
$ (\phi^+,\phi^0)_1 $ & $(1,2,-\frac{1}{2})$ & $\frac{3}{4}(n_1-n_4)$ & + \\
$ (\phi^+,\phi^0)_2 $ & $(1,2,-\frac{1}{2})$& $\frac{1}{4}(9n_1-n_4)$ & + \\
$(\phi^+,\phi^0)_3 $ & $(1,2,-\frac{1}{2})$& $\frac{1}{8}(9n_1-5n_4)$ & - \\
\hline
$ \chi_1 $ & $(1,1,0)$ & $-\frac{1}{2}(3n_1+n_4)$ & + \\
$ \chi_2 $ & $(1,1,0)$ & $-\frac{1}{4}(3n_1+n_4)$ & + \\
$ \chi_3 $ & $(1,1,0)$ & $-\frac{3}{8}(3n_1+n_4)$ & - \\
$ \chi_4 $ & $(1,1,0)$ & $-\frac{3}{4}(3n_1+n_4)$ & + \\
\hline
\end{tabular}
\label{table1}
\end{table}
\end{center}
The third column in table \ref{table1} shows the $U(1)_X$ quantum numbers of various fields which satisfy the anomaly matching conditions. The Higgs content chosen above is not arbitrary and is needed, which leads to the possibility of radiative neutrino masses in a manner proposed in \cite{Ma:2006km} as well as a remnant $Z_2$ symmetry. Two more singlets $S_{1R}, S_{2R}$ are required to be present to satisfy the anomaly matching conditions. In this model, the quarks couple to $\Phi_1$ and charged leptons to $\Phi_2$ whereas $(\nu, e)_L$ couples to $N_R, \Sigma_R$ through $\Phi_3$ and to $S_{1R}$ through $\Phi_1$. The extra four singlet scalars $\chi$ are needed to make sure that all the particles in the model acquire mass. The lagrangian which can be constructed from the above particle content has an automatic $Z_2$ symmetry and hence provides a cold dark matter candidate in terms of the lightest odd particle under this $Z_2$ symmetry.

Let us denote the vacuum expectation values (vev) of various Higgs fields as $ \langle \phi^0_{1,2} \rangle = v_{1,2}, \; \langle \chi^0_{1,2,4} \rangle  =u_{1,2,4}$. We also denote the coupling constants of $SU(2)_L, U(1)_Y, U(1)_X$ as $g_2, g_1, g_x$ respectively. The charged weak bosons acquire mass $M^2_W = \frac{g^2_2}{2}(v^2_1+v^2_2) $. The neutral gauge boson masses in the $(W^{\mu}_3, Y^{\mu}, X^{\mu})$ basis is 
\begin{equation}
M =\frac{1}{2}
\left(\begin{array}{cccc}
\ g^2_2(v^2_1+v^2_2) & g_1g_2(v^2_1+v^2_2) &  M^2_{WX} \\
\ g_1g_2(v^2_1+v^2_2) & g^2_1(v^2_1+v^2_2) & M^2_{YX} \\
\ M^2_{WX} & M^2_{YX}  & M^2_{XX}
\end{array}\right)
\end{equation}
where 
$$M^2_{WX} = -g_2g_x(\frac{3}{4}(n_1-n_4)v^2_1+\frac{1}{4}(9n_1-n_4)v^2_2) $$
$$ M^2_{YX} = -g_1g_x(\frac{3}{4}(n_1-n_4)v^2_1+\frac{1}{4}(9n_1-n_4)v^2_2)$$
$$ M^2_{XX} = g^2_x(\frac{9}{4}(n_1-n_4)^2v^2_1+\frac{1}{4}(9n_1-n_4)^2v^2_2+\frac{1}{16}(3n_1+n_4)^2(4u^2_1+u^2_2+9u^2_4)) $$
The mixing between the electroweak gauge bosons and the additional $U(1)_X$ boson as evident from the above mass matrix should be very tiny so as to be in agreement with electroweak precision measurements. To avoid stringent constraint on mixing here we assume a very simplified framework where there is no mixing between the electroweak gauge bosons and the extra $U(1)_X$ boson. Therefore $ M^2_{WX} = M^2_{YX} = 0$ which gives rise to the following constraint
\begin{equation}
3(n_4-n_1)v^2_1 = (9n_1-n_4)v^2_2
\label{zeromixeq}
\end{equation}
which implies $1 < n_4/n_1 <9 $. If $U(1)_X$ boson is observed at LHC this ratio $n_4/n_1$ could be found empirically 
from its' decay to $q\bar{q}$, $l\bar{l}$ and $\nu\bar{\nu}$ \cite{Adhikari:2008uc}. Here, $q$, $l$ and $\nu$ correspond to 
quarks, charged leptons and neutrinos respectively.
In terms of the charged weak boson mass, we have 
$$ v^2_1 = \frac{M^2_W(9n_1-n_4)}{g^2_2(3n_1+n_4)}, \quad v^2_2 = \frac{M^2_W(-3n_1+3n_4)}{g^2_2(3n_1+n_4)} $$
Assuming zero mixing, the neutral gauge bosons of the Standard Model have masses
$$ M_B = 0, \quad M^2_Z = \frac{(g^2_1+g^2_2)M^2_W}{g^2_2} $$
which corresponds to the photon and weak Z boson respectively. The $U(1)_X$ gauge boson mass is 
$$ M^2_X = 2g^2_X (-\frac{3M^2_W}{8g^2_2}(9n_1-n_4)(n_1-n_4)+\frac{1}{16}(3n_1+n_4)^2(4u^2_1+u^2_2+9u^2_4)) $$

\section{Fermionic Dark Matter Candidates}
\label{fermDM}
In this section we discuss the fermionic dark matter candidates, namely Majorana singlet $N_R$ and triplet $\Sigma_{1R, 2R}$ fermions in the model discussed above. For the computation of the relic density we adopt the simplified analysis done in \cite{Beltran:2008xg}. The relic abundance of a dark matter particle $\chi$ is given by the the Boltzmann equation
\begin{equation}
\frac{dn_{\chi}}{dt}+3Hn_{\chi} = -\langle \sigma v \rangle (n^2_{\chi} -(n^{eqb}_{\chi})^2)
\end{equation}
where $n_{\chi}$ is the number density of the dark matter particle $\chi$ and $n^{eqb}_{\chi}$ is the number density when $\chi$ was in thermal equilibrium. $H$ is the Hubble rate and $ \langle \sigma v \rangle $ is the thermally averaged annihilation cross section of the dark matter particle $\chi$. In terms of partial wave expansion $ \langle \sigma v \rangle = a +b v^2$ the numerical solution of the Boltzmann equation above gives \cite{Kolb:1990vq}
\begin{equation}
\Omega_{\chi} h^2 \approx \frac{1.04 \times 10^9 x_F}{M_{Pl} \sqrt{g_*} (a+3b/x_F)}
\end{equation}
where $x_F = m_{\chi}/T_F$, $T_F$ is the freeze-out temperature, $g_*$ is the number of relativistic degrees of freedom at the time of freeze-out. Dark matter particles with electroweak scale mass and couplings freeze out at temperatures approximately in the range $x_F \approx 20-30$.

The singlet Majorana fermion $N_R$ can be a dark matter candidate if it is the lightest among the $Z_2$-odd particles in the model. To calculate the relic density of $N_R$, we need to find out  its annihilation cross-section to standard model particles. For zero $Z-X$ mixing, the dominant annihilation channel is the one with $X$ boson mediation. The Majorana fermion annihilation with s-channel gauge boson mediation was calculated in \cite{Dreiner:2008tw} and here we follow their approach to find out the annihilation cross section of Majorana fermion $N_R$. 

\begin{figure}[htb]
\centering
\includegraphics{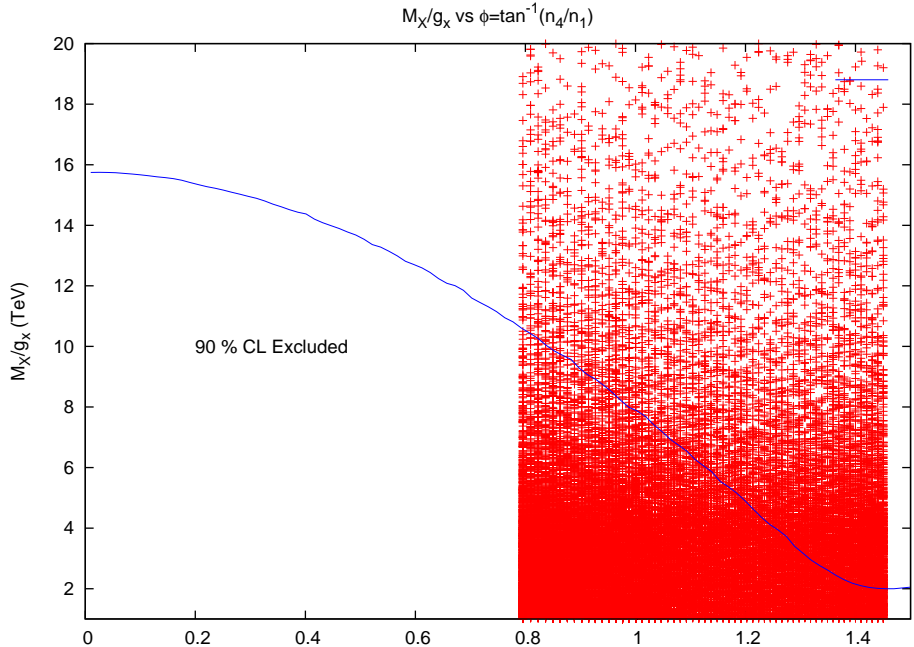}
\caption{Fermion Singlet Dark Matter: Points in the $M_X/g_x-\phi$ plane which satisfy the WMAP dark matter relic density bound $\Omega h^2 \in(0.085,0.139)$ at $3\sigma$ ,the constraints on dark matter-nucleon spin dependent cross section from XENON10 experiment and also the zero-mixing condition (\ref{zeromixeq}). The solid line is the $90 \%$ CL exclusion limit imported from \cite{Adhikari:2008uc}.}
\label{fig1}
\end{figure}

We find out the relic density for a specific choice of $ n_1 $  where we use the normalization $n^2_1 +n^2_4 = 1$. Using the same normalization, the $90\%$ confidence level exclusion on $M_X/g_x$ was shown in \cite{Adhikari:2008uc} which we have shown in figure \ref{fig1} by the solid line. Figure \ref{fig1} also shows the allowed regions of $\phi = \tan^{-1} (n_4/n_1)$ so as the satisfy the zero mixing condition (\ref{zeromixeq}). The points in the plot are those which satisfy the dark matter relic density and direct detection constraints. It can be observed that for all the allowed ranges of $\phi$ we can have observed dark matter relic density. However, the most interesting region of parameter space is near $\phi = 1.5$ where $M_X/g_x$ can be as low as $~2 \; \text{TeV}$ and hence can be probed in the collider experiments.

\begin{figure}[htb]
\centering
\includegraphics{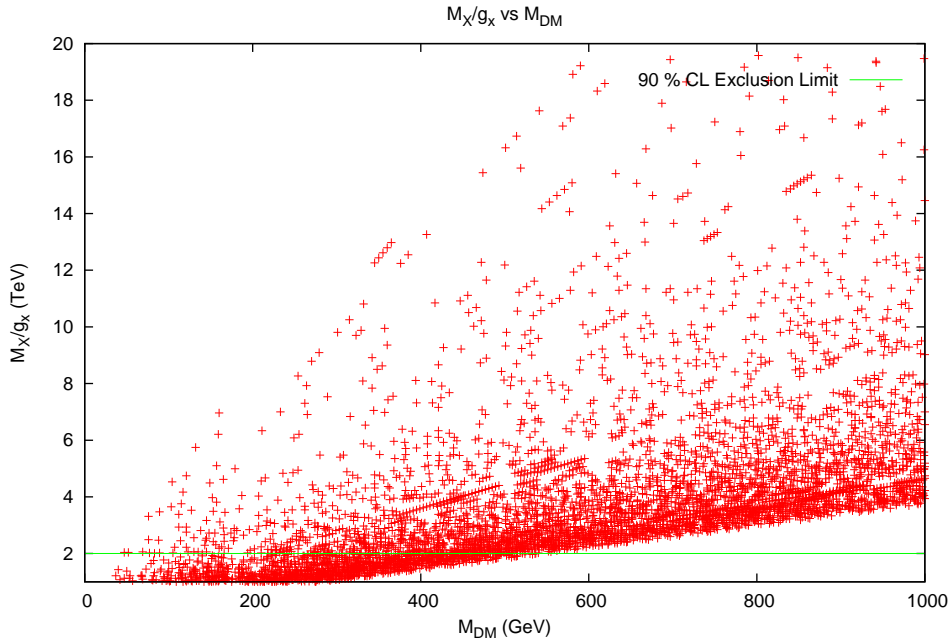}
\caption{Fermion Singlet Dark Matter: Points in the $M_X/g_x-M_{DM}$ plane for $ \phi = \tan^{-1}(n_4/n_1) = 1.45$ which satisfy the WMAP dark matter relic density bound $\Omega h^2 \in(0.085,0.139)$ at $3\sigma$ and the bounds on dark matter-nucleon spin dependent cross section from XENON10 experiment.}
\label{fig2}
\end{figure}

Due to the Majorana nature of the dark matter candidate $N_R$ in the model, there is no spin independent dark matter neucleon scattering mediated by gauge bosons. The only gauge boson mediated scattering is the spin dependent one. 
We follow the analysis of \cite{Jungman:1995df} to find out the spin dependent dark matter-nucleon scattering cross section and use the numerical values of nucleon spin fraction carried by the quarks as given in \cite{Cohen:2010gj}. 
It should be noted that spin independent scattering has very tight constraints coming from null results of CDMSII \cite{Ahmed:2009zw} and XENON100 \cite{Aprile:2010um} whereas constraints on spin dependent scattering are around six  orders magnitudes weaker. We consider the constraints put by XENON10 \cite{Angle:2008we} on spin dependent scattering cross section in the plots \ref{fig1} and \ref{fig2}. Since there is no Higgs in the model which can mediate interaction between $N_R$ and nucleons, our considered model trivially satisfies the strong bounds coming from direct detection experiments on spin independent scattering cross section.

It should be noted that the neutral component of fermion triplet which is odd under the remnant $Z_2$ symmetry can also be a viable dark matter candidate if it is the lightest $Z_2$-odd particle. Such fermion triplet dark matter scenario was studied in \cite{Ma:2008cu} and the authors found that the neutral component of such a fermion triplet has to be in the mass range $ 2.28-2.42 \; \text{TeV} $ to satisfy the relic density bounds. We do not attempt to redo the same analysis here.

\section{Scalar Dark Matter}
\label{scalDM}
Similar to the fermionic dark matter discussed above, the lightest $Z_2$-odd and electrically neutral scalar can also be a viable dark matter candidate. In the model we are studying, there are two such scalar dark matter candidates: the doublet $\Phi_3$ and the singlet $\chi_3$. We discuss both the cases below.

\subsection{Scalar Doublet Dark Matter}
Scalar doublet dark matter has been extensively studied for the last few years. Please see \cite{Barbieri:2006dq,Cirelli:2005uq,LopezHonorez:2006gr} and references therein for earlier works. Generic scalar doublet dark matter models or so called Inert Doublet Models (IDM) contain an additional Higgs doublet odd under a $Z_2$ symmetry which survives even after the electroweak symmetry breaking thereby guaranteeing a stable dark matter candidate. Our considered model not only provides such a $Z_2$-odd Higgs doublet $\phi_3$, but also naturally explains the origin of such a $Z_2$ symmetry. 

The lighter of the electrically neutral CP-odd or CP-even component of such a Higgs doublet $\phi_3$ is the dark matter candidate in such models. As noted in \cite{Adhikari:2008uc}, if the mass difference between the CP-odd and CP-even component are larger than around $1 \; \text{MeV}$, the $Z$ boson mediated scattering with the nucleon where the dark matter particle goes to the next lightest scalar is kinematically disallowed. Such mass splitting can arise from a term $\lambda [(\phi^{\dagger}\phi_3)^2 + \text{h.c.}]$ where $\phi$ is some other Higgs field in the model. However, the gauge structure of the model prevents such a term involving $\phi_3$ and any of the other Higgs fields. As pointed out in \cite{Adhikari:2008uc}, such mass splitting can arise in this model due to the off diagonal elements in the Higgs mass matrix. We assume, throughout our analysis that such splittings are much larger than $1\; \text{MeV}$ required to avoid being ruled out by direct detection experiments like XENON100 \cite{Aprile:2010um}.

We denote the neutral component of $\phi_3$ as $\phi^0_3 = \phi^0_{3R} + i\phi^0_{3I} $ and assume $\phi^0_{3R}$ to be lighter than $\phi^0_{3I}$. This is just a convenient choice and our conclusions remain the same no matter which of $\phi^0_{3R}$ and $\phi^0_{3I}$ is lighter. For relic density calculations, we identify the important annihilation channels for various mass regimes of $\phi^0_{3R}$. For masses below the electroweak boson masses, the annihilation to fermion antifermion pairs via Higgs boson exchange $(\phi^0_{3R} \phi^0_{3R} \rightarrow f \bar{f})$ where $f$ can be any standard model charged fermion except the top quark. Beyond the weak boson threshold the annihilation channel $(\phi^0_{3R} \phi^0_{3R} \rightarrow W^+(Z) W^-(Z))$ opens up and finally after the top quark mass threshold, the annihilation channel into $t \bar{t}$ pair also opens up. We consider the extra $U(1)_X$ boson to be much heavier such that within the region of dark matter mass we study, the annihilation channel into $X$ bosons are not dominant. However, as we will see shortly, the $U(1)_X$ gauge charges affect the relic density due the relation $(\ref{zeromixeq})$ which we choose for zero mixing of the electroweak bosons with the $U(1)_X$ boson.

\begin{figure}[htb]
\centering
\includegraphics{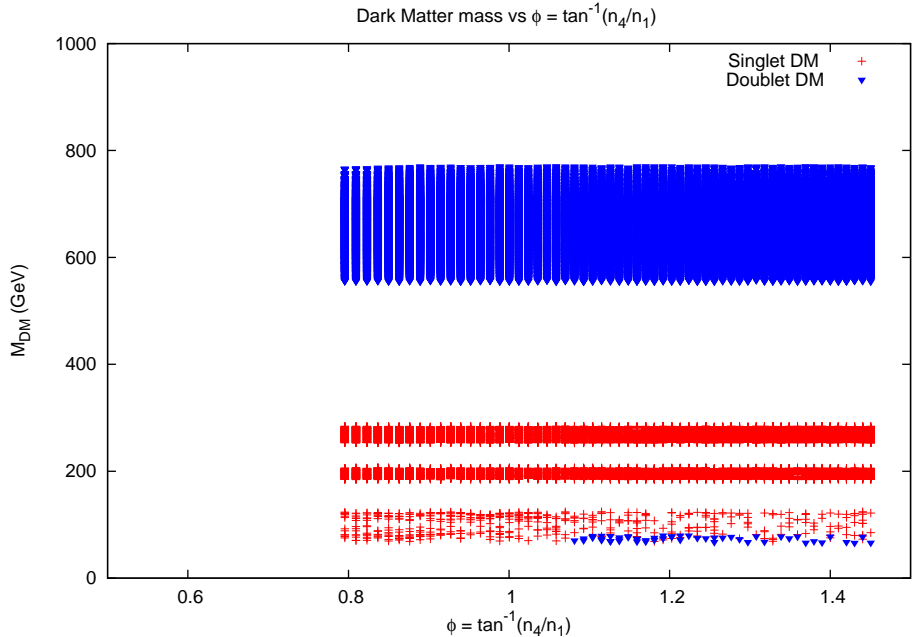}
\caption{Scalar Dark Matter: Points in the $M_{DM}-\phi$ plane which satisfy the WMAP dark matter relic density bound $\Omega h^2 \in(0.085,0.139)$ at $3\sigma$ as well as bounds on spin independent dark matter-nucleon scattering cross section from XENON100.}
\label{fig3}
\end{figure}
\begin{figure}[htb]
\centering
\includegraphics{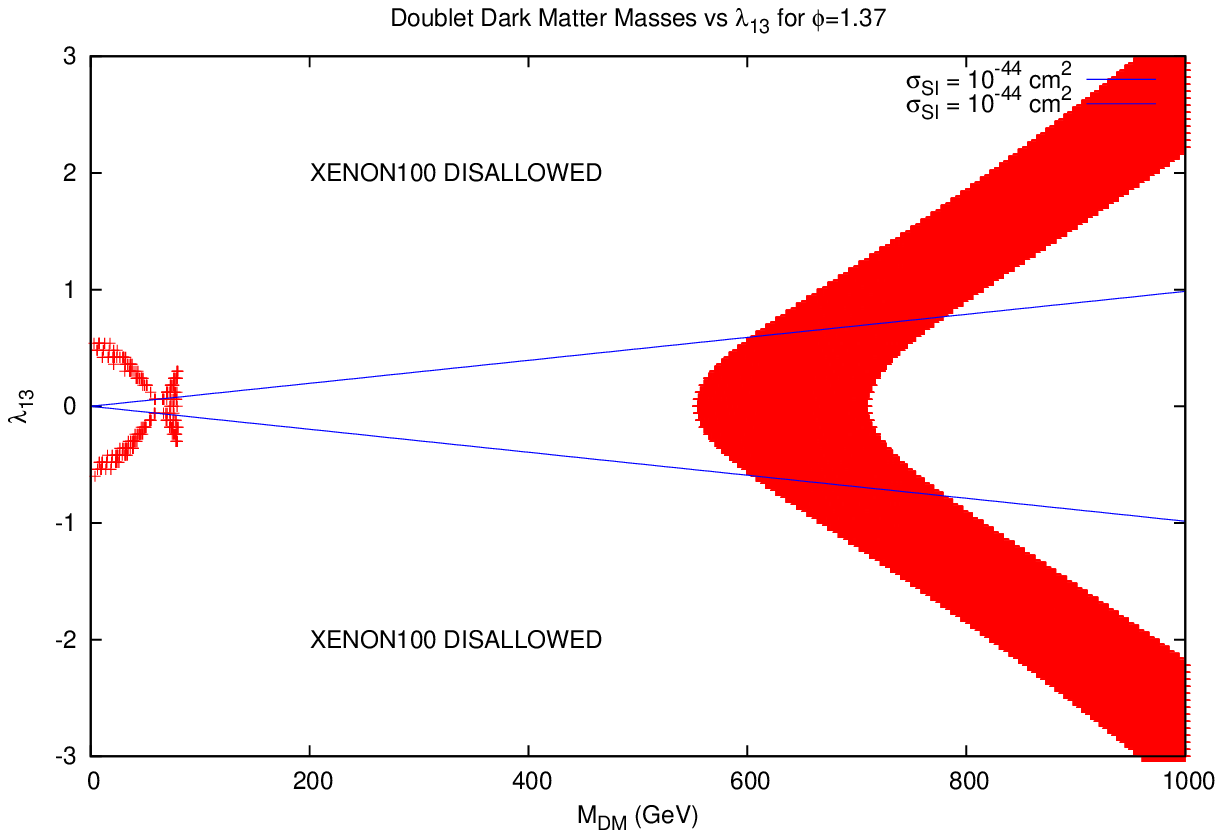}
\caption{Parameter space giving rise to correct WMAP $3\sigma$ relic density is shown in the $\lambda_{13}-M_{DM}$ plane for scalar doublet dark matter. Solid lines refer to the conservative XENON100 upper bound on dark matter nucleon cross-section.}
\label{fig4}
\end{figure}
The relevant part of the scalar potential needed for our calculation is
$$ V_s \supset \mu_1 \chi_1 \chi_2 \chi^{\dagger}_4 + \mu_2 \chi^2_2 \chi^{\dagger}_1 +\mu_3 \chi^2_3 \chi^{\dagger}_4 + \mu_4 \chi_1 \Phi^{\dagger}_1 \Phi_2 + \mu_5 \chi_3 \Phi^{\dagger}_3 \Phi_2 +\lambda_{13} (\Phi^{\dagger}_1 \Phi_1)(\Phi^{\dagger}_3 \Phi_3)$$
$$ +f_1\chi_1\chi^{\dagger}_2\chi^2_3+f_2\chi^3_2\chi^{\dagger}_4+f_3 \chi_1 \chi^{\dagger}_3\Phi^{\dagger}_1\Phi_3 +f_4 \chi^2_2\Phi^{\dagger}_1\Phi_2 + f_5 \chi^{\dagger}_3\chi_4 \Phi^{\dagger}_3 \Phi_2 $$
\begin{equation}
 +\lambda_{23} (\Phi^{\dagger}_2 \Phi_2)(\Phi^{\dagger}_3 \Phi_3)+ \lambda_{16} (\Phi^{\dagger}_1 \Phi_1)(\chi^{\dagger}_3 \chi_3) + \lambda_{26} (\Phi^{\dagger}_2 \Phi_2)(\chi^{\dagger}_3 \chi_3)
\label{scalpot}
\end{equation}
In our considered  model, quarks couple to $\Phi_1$ and charged leptons couple to $\Phi_2$. We take the corresponding couplings and masses into account while calculating the cross section of various annihilation channels. For our calculation we take $m_{\phi_1} = 125 \; \text{GeV} $ and $m_{\phi_2} = 200 \; \text{GeV}$.

For Higgs boson mediated spin independent scattering cross section we follow the analysis of \cite{Barbieri:2006dq} and use the recent lattice calculation values \cite{Giedt:2009mr} of effective matrix elements for Higgs nucleon interactions. We show our results in terms of dark matter mass and $\phi = \tan^{-1}(n_4/n_1)$ in figure \ref{fig3}. Two different dark matter mass ranges are allowed: one in the high mass regime $(\sim 550-750 \; \text{GeV})$ and the other in the low mass regime $(\sim 60-80 \; \text{GeV})$. The allowed region of $\phi$ correspond to the zero mixing condition $(\ref{zeromixeq})$. It is interesting to note that for $\phi$ less than around $1.1$ or so, the low mass regime disappears. Although we are considering a decoupled $U(1)_X$ sector in scalar dark matter analysis, yet the $U(1)_X$ gauge charges affect the parameter space. This is because, the different choices of $n_1$ change the ratio of the vacuum expectation values $v_1$ and $v_2$ by virtue of equation $(\ref{zeromixeq})$. We also show the allowed parameter space in terms of $\lambda_{13}$ (dark matter-Higgs coupling) and dark matter mass giving rise to the correct relic density in figure \ref{fig4} for a particular value of $\phi = \tan^{-1}(n_4/n_1) = 1.37$ . The parameter space
within the two solid lines are allowed from conservative XENON100 upper bound on dark matter nucleon scattering cross section.

\begin{figure}[htb]
\centering
\includegraphics{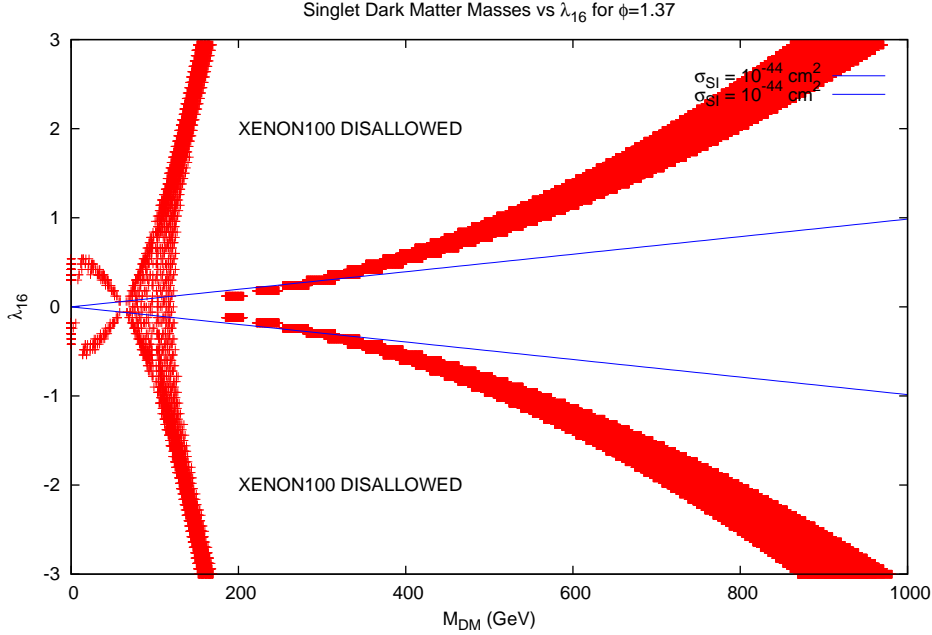}
\caption{Parameter space giving rise to correct WMAP $3\sigma$ relic density is shown in the $\lambda_{16}-M_{DM}$ plane for scalar singlet dark matter. Solid lines refer to the conservative XENON100 upper bound on dark matter nucleon cross-section.}
\label{fig5}
\end{figure}
\subsection{Scalar Singlet Dark Matter}
The analysis for scalar singlet dark matter is similar to the doublet case discussed above except certain differences. We denote the $Z_2$-odd singlet as $\chi_3 = \chi_{3R} + i \chi_{3I}$. As discussed above, in the doublet case the scalar potential does not contain mass splitting terms between CP-even and CP-odd electrically neutral Higgs, and we assume the off diagonal terms in the Higgs mass matrix to generate such splittings. However, in case of the singlet, such terms are present in the scalar potential which naturally generate the desired mass splitting between $\chi_{3R}$ and $\chi_{3I}$. These terms involve the couplings $\mu_3, f_1$ as can be seen in equation $(\ref{scalpot})$. 

Unlike the doublet case, here the dark matter annihilation channels into electroweak gauge bosons are absent. And, we also assume the $U(1)_X$ gauge boson to be heavier than the dark matter mass considered and hence annihilation channel into $X$ boson is also absent. The relevant annihilation channel is $(\chi_{3R} \chi_{3R} \rightarrow f \bar{f})$ where $f$ denotes quarks and charged leptons. The annihilation into quark-antiquark pairs is mediated by $\phi_1$ whereas that to charged lepton-antilepton pairs is mediated by $\phi_2$. We show our results in terms of dark matter mass and $\phi = \tan^{-1}(n_4/n_1)$ in figure \ref{fig3}. In this case, there are three allowed mass ranges: around $100, 200, 250 \; \text{GeV}$ respectively. Unlike in the scalar doublet dark matter case, here the allowed dark matter mass range is independent of $\phi$. In other words for entire allowed range of $\phi$ satisfying the zero mixing condition (\ref{zeromixeq}), both light and heavy dark matter masses are allowed. Similar to the scalar doublet dark matter case, here also we show the allowed parameter space in terms of $\lambda_{16}$ (dark matter-Higgs coupling) and dark matter mass giving rise to the correct relic density in figure \ref{fig5} for a particular value of $\phi = \tan^{-1}(n_4/n_1) = 1.37$.

\subsection{Effect of $U(1)_X$ gauge boson}
So far we have been assuming the $U(1)_X$ boson to be heavier than $1 \; \text{TeV}$ and hence not contributing to the dark matter annihilation channels. However, as seen in figure \ref{fig1}, $M_X$ can be lighter than a TeV provided $M_X/g_x > 2 \; \text{TeV}$. This is possible when $\phi = \tan^{-1}(n_4/n_1)$ is close to $\pi/2$. We now consider such a light $M_X$ so that the annihilation channel of scalar dark matter to the gauge boson pairs $X X$ opens up. We take $M_X/g_x > 2 \; \text{TeV}$ and $\phi = \tan^{-1}(n_4/n_1) = 1.45$ for the analysis. We find that this new annihilation channel do not affect the annihilation much, because a light $M_X$ is possible only when the corresponding gauge coupling $g_x$ is tiny so as to satisfy $M_X/g_x > 2 \; \text{TeV}$. 

\section{Neutrino Mass}
\label{sec:numass}

In this section, we summarize the origin of neutrino mass in the model. The relevant part of the Yukawa Lagrangian is 
$$ \mathcal{L}_Y \supset y \bar{L} \Phi^{\dagger}_1 S_{1R} + h_N \bar{L} \Phi^{\dagger}_3 N_R + h_{\Sigma}  \bar{L}\Phi^{\dagger}_3 \Sigma_R + f_N N_R N_R \chi_4+ f_S S_{1R} S_{1R} \chi_1 $$
\begin{equation}
+ f_{\Sigma} \Sigma_R \Sigma_R \chi_4 + f_{NS} N_R S_{2R} \chi^{\dagger}_2 + f_{12} S_{1R} S_{2R} \chi^{\dagger}_3
\label{yukawa} 
\end{equation}

\begin{figure}[htb]
\centering
\includegraphics[scale=0.75]{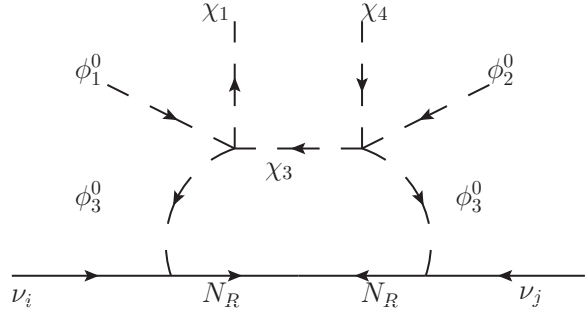}
\caption{One-loop contribution to neutrino mass}
\label{numass}
\end{figure}

The Majorana mass of the fermions $S_R, N_R$ and $\Sigma_R$ arise as a result of the spontaneous symmetry breaking of $U(1)_X$ symmetry by the vev of $\chi_{1,2,4}$. Neutrinos acquire Dirac masses by virtue of their couplings to $S_{1R}$ as shown in equation $(\ref{yukawa})$. Thus the $3 \times 3$ neutrino mass matrix receives tree level contribution from standard type I seesaw mechanism and one gets the hierarchical pattern of neutrino mass with only one massive and other two neutrinos massless. $S_{1R}$ can couple to arbitrary linear combination of $\nu_i$ by assigning different values of $y$ in
equation (\ref{yukawa}) for
different generation.  However, considering it non-zero  and same for $\nu_\mu$ and $\nu_\tau$ only,  the heaviest mass $m_{\nu 3}$ is 
given by 
\begin{equation}
m_{\nu 3} \approx \frac{ 2y^2 v_1^2}{f_S u_1}
\label{neutmass}
\end{equation}
which sets the scale of higher neutrino mass square difference of  about $ 2.4  \times 10^{-3}$ eV$^2$ and for that
$m_{\nu 3}  $ may be considered to be about $0.05$ eV (for hierarchical neutrino masses) to about 0.1 eV (for almoset degenerate neutrino masses).
Two neutrinos which are massless at the tree level become massive from one loop contribution 
as shown in figure \ref{numass} of Feynman diagram involving one of $N_R, \Sigma^0_{1R, 2R}$. These contributions 
set the scale of lower mass squared difference of about $7.6 \times 10^{-5}$ eV$^2$ for which other neutrino masses may be considered to be about $10^{-2}$ eV (for hierarchical neutrino masses ) to about 0.1 eV (for almost degenerate neutrino masses). Somewhat similar to \cite{Ma:2006km}, such a one loop diagram gives partial
contribution through   $A_k$ as mentioned below when there is a mass splitting between the CP-even and CP-odd neutral components of the Higgs field involved in the loop which is $\phi^0_3$ in this case. In our considered model, such a mass splitting is possible due to the  couplings between $\phi^0_3$ and the singlet scalar fields $\chi$ shown in equation $(\ref{scalpot})$. Such a mass splitting is also necessary for $\phi^0_3$ to be a dark matter candidate as discussed above.

The  one-loop contribution $(M_\nu)_{ij}$ to neutrino mass is given by

\begin{eqnarray} 
({M_\nu)}_{ij} \approx  \frac{f_3 f_5 v_1 v_2 u_1 u_4}{16 \pi^2} \sum_k {h_{N, \Sigma} }_{ik} {h_{N, \Sigma} }_{jk} \left( A_k +{(B_k)}_{ij} \right)
\label{nuradmass}
\end{eqnarray}
where 
\begin{eqnarray}
A_k &=& {(M_{N, \Sigma})}_k \left[ I\left( m_{\phi^0_{3R}},m_{\phi^0_{3R}},{(M_{N, \Sigma})}_k, m_{\chi_{3R}} \right) - I\left(m_{\phi^0_{3I}},m_{\phi^0_{3I}},{(M_{N, \Sigma})}_k, m_{\chi_{3R}} \right) \right], 
\end{eqnarray}
\begin{eqnarray}
{(B_k)}_{ij}= -(2-\delta_{ij}) {(M_{N, \Sigma})}_k I\left(m_{\phi^0_{3R}},m_{\phi^0_{3I}},{(M_{N, \Sigma})}_k, m_{\chi_{3I}} \right),
\end{eqnarray}
in which
\begin{eqnarray}
I(a,a,b,c)=  \frac{(a^4-b^2 c^2) \ln (a^2/c^2)}{{(b^2 -a^2)}^2 {(c^2 -a^2)}^2}+ \frac{b^2 \ln (b^2/c^2)}{ (c^2 - b^2){(a^2 - b^2)}^2}  
  -\frac{1}{(a^2 - b^2) (a^2 -c^2)},
\end{eqnarray}
\begin{eqnarray} 
I(a,b,c,d)&=&  \frac{1}{a^2-b^2}  \left[  \frac{1}{a^2-c^2} \left( \frac{a^2}{a^2-d^2} \ln(a^2/d^2) - \frac{c^2}{c^2-d^2} \ln(c^2/d^2)\right)\right. \nonumber \\  &-&  \left. \frac{1}{b^2-c^2} \left( \frac{b^2}{b^2-d^2} \ln(b^2/d^2) - \frac{c^2}{c^2-d^2} \ln(c^2/d^2)\right)  \right]
\end{eqnarray}
and
$m_{\phi^0_{3R}}$ and $m_{\phi^0_{3I}} $ are the masses corresponding to $Re[\phi^0_3]$ and $Im[\phi^0_3]$ respectively
and  $m_{\chi_{3R}}$ and  $m_{\chi_{3I}}$ are the masses corresponding to $Re[\chi^0_3]$ and $Im[\chi^0_3]$ respectively. We have considered three types of dark matter and for those $A_k$ and $B_k$ may be approximated as given below. For that we
define $\delta m^2 ={m_{\phi^0_{3I}}}^2-{m_{\phi^0_{3R}}}^2 $ and $  m_0^2 = \left( {m_{\phi^0_{3R}}}^2+
{m_{\phi^0_{3I}}}^2 \right)/2 $ and $(M_{N, \Sigma})_k = m_{2k}$, $m_{\chi_{3R}} = m_3$ and $m_{\chi_{3I}} = m_4$. We have neglected the mixing between $\phi_3^0$ and $\chi_3^0$.

\noindent
(a) For fermionic dark matter one of the masses $(M_{N, \Sigma})_k$ is lightest and if we assume
$(M_{N, \Sigma})_k << m_{\phi^0_{3R}}, m_{\phi^0_{3I}}, m_{\chi_{3R}},  m_{\chi_{3I}}$ then
\begin{eqnarray}
A_k \approx  \frac{\delta m^2  m_{2k}}{m_0^4} \left[\frac{2 m_{2k}^2 \ln (m_0^2/m_3^2)  }{\left(m_0^2 -m_3^2\right)^2}
-\frac{m_{2k}^2 \ln(m_{2k}^2/m_3^2) }{m_3^2 m_0^2} + \frac{1}{\left( m_0^2-m_3^2 \right) } \right]\; .
\label{fermA}
\end{eqnarray}
(b) For scalar doublet  dark matter one of the masses $m_{\phi^0_{3R}}$ and $m_{\phi^0_{3I}}$ is lightest and if we assume
$m_{\phi^0_{3R}}, m_{\phi^0_{3I}} << (M_{N, \Sigma})_k, m_{\chi_{3R}},  m_{\chi_{3I}}$ then
\begin{eqnarray}
A_k \approx  \frac{\delta m^2 m_{2k} }{m_{2k}^4 m_3^4}  \left[    (m_{2k}^2 + m_3^2 ) \left\{2  \ln (m_0^2/m_3^2) -1  \right\}
-\frac{ m_3^4 \ln (m_{2k}^2/m_3^2) }{(m_{2k}^2 -m_3^2) } \right] \;
\label{scaldA}  .
\end{eqnarray}
(c) For scalar singlet dark matter  if we consider $  m_{\chi_{3R}} = m_{\chi_{3I}} $ and 
$m_{\chi_{3R}} << (M_{N, \Sigma})_k , m_{\phi^0_{3R}}, m_{\phi^0_{3I}}$  then
\begin{eqnarray}
A_k \approx    \frac{ \delta m^2 m_{2k} }{ m_0^4 } \left[ \frac{ 2 m_3^2 \ln (m_0^2/m_3^2) }{{(m_0^2-m_{2k}^2)}^2}  + 
\frac{1}{(m_0^2 -m_{2k}^2 )}   \right].
\label{scalsA}
\end{eqnarray}
For all the above three cases $(B_k)_{ij}$ can be approximated as
\begin{eqnarray}
(B_k)_{ij} \approx -  \frac{(2-\delta_{ij}) \; m_{2k} \ln \left(m_0^2/m_4^2  \right) }{\left(  m_0^2 - m_{2k}^2  \right)\left( m_0^2 -m_4^2 \right)}.
\label{Ball}
\end{eqnarray}
If all the scalar masses in the loop diagram are almost degenerate and written as $m_{sc}$ then 
\begin{eqnarray}
A_k + (B_k)_{ij} \approx m_{2k} \left[\frac{m_{sc}^2 
+ m_{2k}^2 }{m_{sc}^2 \left( m_{sc}^2 - m_{2k}^2 \right)^2 }- \frac{(2-\delta_{ij})\; m_{2k}^2}{\left(m_{sc}^2 - m_{2k}^2 \right)^3}\ln \left( m_{sc}^2/m_{2k}^2 \right)   \right],
\label{scaldeg}
\end{eqnarray}
and if all scalar and fermion masses in the loop are almost degenerate and written as $m_{deg}$ then
\begin{eqnarray}
A_k + (B_k)_{ij} \approx  \frac{(2-\delta_{ij})}{6 m_{deg}^3}\; .
\label{fermscal}
\end{eqnarray}
$M_{N,\Sigma}$ is the Majorana mass term of $N_R(\Sigma^0_R)$. $h_{N, \Sigma}$ are the Yukawa couplings in equation $(\ref{yukawa})$. 
To get appropriate neutrino mass square differences we shall require these loop contributions to be about $10^{-2}$ eV 
(for hierarchical neutrino masses) to about $0.1$ eV {for almost degenerate neutrino masses) as mentioned earlier. 

To check whether the allowed dark matter masses can give rise to the observed neutrino mass, we consider the simplest case as an example where the all the scalar masses in the loop are almost degenerate (\ref{scaldeg}). We take $v_1 = v_2 \sim 10^2 \; \text{GeV}$ and $u_1 = u_2 \sim M_X/g_x \geq 2 \; \text{TeV}$. For singlet fermion $N_R$ as light as $50 \; \text{GeV}$ (which can satisfy dark matter constraints as discussed in section \ref{fermDM}) and assuming the scalar masses involved in (\ref{scaldeg}) to be of order $100 \; \text{GeV}$, the expression of neutrino mass (\ref{nuradmass}) forces one have the product of some Yukawa couplings as $f_3f_5 h^2_{N, \Sigma} \leq 10^{-9} $ to give rise to neutrino mass of order $10^{-2} \; \text{eV}$ for hierarchical neutrino masses. This product is $\leq 10^{-10}$ for almost degenerate neutrino masses. However, presently neutrinos are supposed to be not a strong candidate for hot dark matter and one may refrain from considering degenerate mass patttern.  This is not a severe fine-tuning of the Yukawa couplings looking at the fact that electron Yukawa coupling in the Standard Model is of order $10^{-5}$. 

For scalar doublet dark matter we have $\delta m^2 \geq 1 \; \text{MeV}^2$ which is required from dark matter direct detection constraints, but the mass square differences of other scalar and fermion in the loop are much larger than $\delta m^2$ and  $A_k$ is  smaller than $(B_k)_{ij}$ in general case. For both the general case (\ref{Ball}) as well as the simpler case (\ref{scaldeg}), we require more fine-tuning of the parameters $f_3, f_5, h_{N, \Sigma}$ in neutrino mass expression (\ref{nuradmass}). This is because of the presence of singlet fermion mass in the numerator which is heavier in the scalar dark matter case than in the fermion dark matter case. We leave a more general study of neutrino mass in this model for future work.

\section{Implications on LHC Higgs Search}
\label{sec:lhc}

For the last one year or so, LHC has been performing extremely well and has brought down the SM like Higgs mass range to a small window $115-127 \; \text{GeV}$ by excluding the mass range $127-600 \; \text{GeV}$ at $95 \%$ confidence level \cite{lhc:2011}. The same results also indicate slight excess of events in the Higgs decaying to diphoton channel which correspond to a Higgs with a mass of around $125 \; \text{GeV}$. If a Higgs boson with mass $125 \; \text{GeV}$ is confirmed by LHC in near future, then such a Higgs will correspond to $\phi^0_1$ in our model which couples to quarks and hence produced dominantly at LHC through gluon gluon fusion. However, the Higgs coupling to leptons ($\phi^0_2$ in our model) will be produced much inefficiently and hence can well be in LHC exclusion range for SM like Higgs boson. We have assumed the mass of $\phi^0_2$ to be $200 \; \text{GeV}$. If LHC confirms a SM like Higgs with mass $M_H = 125 \; \text{GeV}$ it can have important implications for scalar dark matter models such as ruling out scalar dark matter with mass below $M_H/2 \sim 60 \; \text{GeV}$  \cite{Djouadi:2011aa} due to the large invisible Higgs decay width.

However, if LHC does not confirm a SM like light Higgs, still there can be a light Higgs like $\phi^0_2$ in our model which couple to leptons only and hence not produced dominantly at LHC. In such a scenario, a light scalar Dark matter will still be allowed whose annihilation channels will be dominated by $\phi^0_2$ mediated ones to lepton-antilepton pairs. Also, if LHC closes the existing window of $115-127 \; \text{GeV}$ SM like Higgs, then the Higgs coupling to quarks in our model should be heavier than the LHC search range i.e. close to the unitarity bound $\sim 700 \; \text{GeV}$ \cite{unitarity}. Such a scenario will also ease the tension between fourth generation models and LHC Higgs exclusion range $\sim 120-600 \; \text{GeV}$ at $95\%$ C.L \cite{Koryton2011} for SM with four generations. Recently, one of us (DB) proposed another mechanism to ease this tension by an abelian gauge extended model where the fourth generation couples to a heavier Higgs whereas the first three generations couple to a SM like light Higgs \cite{Borah:2011ve}. The present model (after including the fourth family) is different in the sense that here all the four generation quarks couple to a heavy Higgs and all the four generation leptons couple to a light Higgs. However this possibility will be ruled out if LHC confirms a SM like light Higgs. Thus our model provides various interesting possibilities which will either be confirmed or ruled out very soon by the LHC experiment.

\section{Results and Conclusion}
\label{results}
In this work, we have discussed such an abelian gauge extension of standard model originally proposed by \cite{Adhikari:2008uc} in details. We study the $U(1)_X$ as well as the electroweak symmetry breaking in this model giving rise to the observed gauge bosons and a heavy $X$ boson beyond the LHC search range so far. The residual $Z_2$ symmetry of the model makes it very rich phenomenologically. The lightest $Z_2$-odd particle is naturally stable in this model and hence can be a good dark matter candidate provided other relevant constraints from cosmology and direct detection experiments are satisfied. For simplicity, we assume zero tree level mixing of this extra gauge boson with the observed electroweak bosons which gives rise to a constraint $(\ref{zeromixeq})$ relating the vev's of $SU(2)_L$ doublet Higgs fields and $U(1)_X$ gauge charges of various fields in the model. 

 We have studied the fermionic dark matter candidates in section \ref{fermDM}. The singlet fermion $N_R$ can be as light as around $50 \; \text{GeV}$ for certain specific choices of $U(1)_X$ gauge charges and couplings. The triplet fermionic dark matter, on the other hand has to be heavier than $2.28 \; \text{TeV}$ to satisfy the relic density bounds as discussed in \cite{Ma:2008cu}. We have not repeated the same discussion of triplet dark matter in our work.

We discuss both scalar doublet and scalar singlet dark matter candidates in our model. We find that in case of scalar doublet dark matter, both relic density and direct detection bounds can be satisfied for a wide range of dark matter masses. The allowed mass falls into two regimes: one in the $(\sim 60-80 \; \text{GeV})$ range and the other in $(\sim 550-750 \; \text{GeV})$ range as shown in figure \ref{fig3}. Interestingly, we find that for $\phi = \tan^{-1}(n_4/n_1) $ below $1.1$ or so, the low mass regime of dark matter completely disappears. The allowed mass range for scalar singlet dark matter, as seen in figure \ref{fig3}, is either $(\sim 60-130 \; \text{GeV})$ or $200-300 \; \text{GeV}$. Unlike in the doublet case, here different choices of $\phi = \tan^{-1}(n_4/n_1)$ do not affect the allowed dark matter mass range. We also check that the inclusion of scalar dark matter annihilation into $U(1)_X$ gauge boson pairs do not affect the allowed dark matter mass range.

The neutrino mass can arise at both tree level as well as one loop level in this model. Tree level origin of neutrino mass is the standard type I seesaw mechanism whereas the radiative origin involves the $Z_2$-odd fields (both scalar and fermionic) in the loop. Non-zero loop contribution is possible only when there is a mass splitting between the real and imaginary components of the Higgs fields involved in the loop. Such mass splitting trivially arises in the scalar potential for the singlet Higgs field, whereas for the doublet Higgs field such splitting arises through the off-diagonal terms in the mass matrix. Such mass splitting is also necessary for the $Z_2$-odd scalar to be a dark matter candidate so as to avoid being ruled out by direct detection experiments. We show that the allowed mass ranges for various dark matter candidates in this model are also consistent with the observed neutrino mass of the order $0.1 \; \text{eV}$. We point out the interesting possibilities our model offer in terms of LHC Higgs search results both for usual three generation SM as well four generation SM. These possibilities will either be ruled out or confirmed soon by the LHC.

\section{Acknowledgement}
DB would like to acknowledge the hospitality at CTP-JMI, New Delhi during the 2011 SERC THEP Main school where part of this work was completed.


\end{document}